\documentclass[a4paper,11pt]{article}
\usepackage{amsmath,amssymb,graphicx}
\usepackage{cite}
\usepackage{array,multirow}
\usepackage{multicol,color,longtable}
\usepackage[hyphens,spaces,obeyspaces]{url}
\usepackage{hyperref}
\definecolor{darkred}{rgb}{0.65,0.15,0}
\hypersetup{pdfborder={0 0 0},colorlinks=true,urlcolor=blue,citecolor=blue,linkcolor=darkred,linktocpage=true}

\usepackage{young}
\usepackage[vcentermath]{youngtab}

\newcommand{\eprint}[1]{{\href{http://arxiv.org/abs/#1}{[\texttt{#1}]}}}
\newcommand{\eprintN}[1]{{\href{http://arxiv.org/abs/#1}{[\texttt{#1 [hep-th]}]}}}
\newcommand{\eprintQC}[1]{{\href{http://arxiv.org/abs/#1}{[\texttt{#1 [gr-qc]}]}}}

\newcommand{\eprintCO}[1]{{\href{http://arxiv.org/abs/#1}{[\texttt{#1 [astro-ph.CO]}]}}}
\newcommand{\doi}[2]{{\hypersetup{urlcolor=darkred}\href{http://dx.doi.org/#2}{#1}\hypersetup{urlcolor=blue}}}

\newcommand{\nn}{\nonumber}
\newcommand{\mf}[1]{{\mathfrak{#1}}}

\newcommand{\ints}{\mathbb{Z}}

\newcommand{\cx}{\mathbb{C}}

\newcommand{\LL}{R}

\newcommand{\lb}{\left[}
\newcommand{\rb}{\right]}

\newcommand{\Poincare}{Poincar\'{e}}

\newcommand{\Minfty}{\text{Maxwell}_\infty}
\newcommand{\Pinfty}{\text{\Poincare}_\infty}
\newcommand{\AdSinfty}{\text{Poincar\'e}_\infty}
\newcommand{\STinfty}{\mathcal{M}_\infty}

\makeatletter

\@addtoreset{equation}{section} \makeatother

\begin{document}
\setcounter{page}{0}

{\flushright {ICCUB-20-013}\\[10mm]}

\begin{center}
{\LARGE \bf A free Lie algebra approach to\\[2mm]
 curvature corrections to flat space-time\\[10mm]}

\vspace{8mm}
\normalsize
{\large  Joaquim Gomis${}^{1}$, Axel Kleinschmidt${}^{2,3}$, Diederik Roest${}^4$ and Patricio Salgado-Rebolledo${}^5$}

\vspace{10mm}
${}^1${\it Departament de F\'isica Qu\`antica i Astrof\'isica\\ and
Institut de Ci\`encies del Cosmos (ICCUB), Universitat de Barcelona\\ Mart\'i i Franqu\`es, ES-08028 Barcelona, Spain}
\vskip 1 em
${}^2${\it Max-Planck-Institut f\"{u}r Gravitationsphysik (Albert-Einstein-Institut)\\
Am M\"{u}hlenberg 1, DE-14476 Potsdam, Germany}
\vskip 1 em
${}^3${\it International Solvay Institutes\\
ULB-Campus Plaine CP231, BE-1050 Brussels, Belgium}
\vskip 1 em
${}^4${\it Van Swinderen Institute for Particle Physics and Gravity,
University of Groningen, Nijenborgh 4, 9747 AG Groningen, The Netherlands}
\vskip 1 em
${}^5${\it School of Physics and Astronomy, University of Leeds\\ Leeds, LS2 9JT, United Kingdom}
\vspace{15mm}

\hrule

\vspace{2mm}

\begin{tabular}{p{13cm}}
{\small
We investigate a systematic approach to include curvature corrections to the isometry algebra of flat space-time order-by-order in the curvature scale. The \Poincare~algebra is extended to a free Lie algebra, with generalised boosts and translations that no longer commute. The additional generators satisfy a level-ordering and encode the curvature corrections at that order. This eventually results in an infinite-dimensional algebra that we refer to as $\Pinfty$, and we show that it contains among others an (A)dS quotient. We discuss a non-linear realisation of this infinite-dimensional algebra, and construct a particle action based on it. The latter yields a geodesic equation that includes (A)dS curvature corrections at every order. }
\end{tabular}
\vspace{3mm}
\hrule
\end{center}

\thispagestyle{empty}

\newpage

\setcounter{page}{1}
\setcounter{tocdepth}{2}
\tableofcontents

\vspace{5mm}
\hrule
\vspace{5mm}

\section{Introduction}\label{Introduction}

The laws of elementary particle physics are relativistic to very high precision.
This is described by Minkowski geometry whose isometries span the \Poincare~algebra that provides the underlying kinematical
structure for the vast majority of field theories. However, in a gravitational context,  
Minkowski space is replaced by a curved space-time\footnote{Whether or not the mere notion of space-time remains appropriate at very high energies and curvatures is open.} and 
  a range of astronomical and cosmological observations indicate that we live in an expanding  and accelerating space-time. To very good accuracy (see e.g.~\cite{Aghanim:2018eyx}, and modulo the currently emerging Hubble tension), its evolution is described by $\Lambda$-CDM, whose dominant component is a very small and positive cosmological constant (of the meV order). In the absence of matter 
  (which will be redshifted in the future and become more and more negligible), the resulting space-time would be de Sitter. 
  
The smallness of the cosmological constant in Planck units implies that the de Sitter isometries are a small perturbation of the \Poincare~ones. In this paper, we aim to investigate such corrections to Poincar\'e isometries: what structures do these give rise to, what are symmetries of these structures and can we systematically describe these? 

The hallmark of a curved space-time is that covariant derivatives no longer commute. In the case of de Sitter this is expressed in the global isometry algebra by
\begin{align}
\label{eq;AdSintro}
[P_a, P_b] = 
\frac{\sigma}{\LL^2} J_{ab}\,,
\end{align}
where translations generate Lorentz transformations. In the above relation, we have introduced a parameter $\sigma=\pm 1$ that allows us to treat the de Sitter ($\sigma=-1$) and Anti de Sitter case ($\sigma=+1$) on the same footing.\footnote{In the main body of the paper we will use the convention that the translation generators are dimensionless by absorbing the length scale into the generators. The algebra in this convention is given in appendix~\ref{app:AdS}.}

Our investigation parallels to some extent the inclusion of the speed of light corrections in Galilean theories,  see for example
\cite{dautcourt1990newtonian,DePietri:1994je,
Dautcourt:1996pm,VandenBleeken:2017rij,
Hansen:2018ofj,Hansen:2019vqf,Bergshoeff:2019ctr,
Hansen:2019svu,Gomis:2019sqv,Ergen:2020yop}. 
In this (non-)relativistic case we have two distinct ways of understanding the symmetry algebra. Starting from the relativistic \Poincare{} algebra, one can perform an algebra contraction to the non-relativistic Galilei algebra. Instead, starting from the Galilei transformations one can include relativistic corrections at every order in $1/c$. With corrections up to a finite order, this system has neither Galilei nor \Poincare{} symmetry. Only when including an infinite set of corrections, with specific coefficients, does one regain the \Poincare{} symmetry. However, it has recently been argued that the finite-correction-case has a symmetry algebra, but this requires enlarging the space on which the algebra acts. The Galilei algebra 
appears as a simplest quotient of this algebra, while
taking bigger quotients one obtains the non-relativistic algebras studied in~\cite{Hansen:2018ofj,Hansen:2019vqf,Bergshoeff:2019ctr,Ozdemir:2019orp,Gomis:2019sqv,Hansen:2019svu,Gomis:2019nih}.
When including corrections at all orders, this becomes an infinite-dimensional algebra
\cite{Hansen:2019vqf,Bergshoeff:2019ctr,Gomis:2019fdh,Gomis:2019nih}. The \Poincare{} algebra arises as a finite-dimensional quotient of this.

We shall demonstrate that an analogous structure governs the (A)dS
curvature corrections to Poincar\'e. Starting from the commuting translations $P_a$ we build up an algebra structure where the most general commutators allowed by Lie algebra cohomology are included. The resulting free Lie algebra, called $\Minfty$, is infinite-dimensional and has already appeared in~\cite{Bonanos:2008ez,Gomis:2017cmt} in the context of electro-magnetism. We here show how to identify a quotient of this algebra that corresponds to a small curvature expansion of (A)dS space. This quotient is also infinite-dimensional.

The structure is similar to~\cite{Gomis:2019sqv} in that we can describe the same quotient as a Lie algebra expansion~\cite{Hatsuda:2001pp,deAzcarraga:2002xi,Izaurieta:2006zz} of the (A)dS algebra, which can be thought of as systematically including $1/\LL$-corrections where $\LL$ is a generic length that characterises the scale of the curvature, as shown in~\eqref{eq;AdSintro}. In the non-relativistic case the same was achieved  for $1/c$-corrections. For the de Sitter case, the scale $\LL$ here is related to the Hubble parameter of the accelerating Universe.  
Asymptotically in the expansion method, this results in an infinite-dimensional algebra that is an extension of the original \Poincare~starting point. It includes an infinite number of generalised translations and Lorentz transformations. This situation is similar to the non-relativistic results of~\cite{Gomis:2019fdh,Khasanov:2011jr,Hansen:2019vqf}.
Appropriately identifying the generators at different levels gives rise to the (A)dS algebra, which is therefore a quotient of the infinite-dimensional algebra. 

Turning to dynamics, we use the method of non-linear realisation~\cite{Coleman:1969sm,Callan:1969sn,Salam:1969rq,Salam:1970qk,Isham:1971dv,Volkov:1973vd,Ogievetsky:1973ik,Ogievetsky:1974,Borisov:1974bn,Ivanov:1975zq} to construct particle actions governed by the extended \Poincare~algebra $\Pinfty$, and show how these include the effect of curvature order-by-order. Independent of the choice of coefficients, this allows one to construct the (Anti) de Sitter corrections. We show that the particle in this case includes the geodesic equation of (Anti) de Sitter. This demonstrates how the subsequent inclusion of curvature corrections interpolates from the \Poincare{} to (Anti) de Sitter case.

This paper is organised as follows. We first study the most general extension of the Poincar\'e algebra when allowing for non-commuting translations and how they generate the infinite-dimensional algebra $\Minfty$. We then consider a quotient to another infinite-dimensional algebra $\Pinfty$ and show that it agrees with the Lie algebra expansion method applied to the (A)dS algebra, giving the first hint of connections to curved space-time. A certain non-linear realisation of the $\Pinfty$ symmetry is then studied in section~\ref{sec:NRPinfty}, where we also relate the resulting coset space $\STinfty$ to flat space and (A)dS. In section~\ref{sec:part}, we study particle dynamics on $\STinfty$ and show how it can be understood as a systematic $1/\LL$ expansion of the (A)dS geodesic before concluding. Some relevant standard results on (A)dS are collected in an appendix.

\section{The $\Pinfty$ algebra}

In this section we consider the most general extension of the Poincar\'e algebra in order to capture curvature corrections to the flat space-time case. This will lead to an infinite-dimensional algebra with a close relation to the (A)dS algebra. For the benefit of the reader we summarise a number of relevant points of the (A)dS algebra in appendix~\ref{app:AdS}.

\subsection{Free Lie algebras and $\Minfty$}

As mentioned in the introduction, one hallmark of both non-trivial gauge or geometric connections is that the corresponding covariant derivatives no longer commute. This is clear for the case of (A)dS space where one still has global isometries but now the corresponding translations are non-commuting, see~\eqref{eq:AdS}. In contrast, the isometries of the flat space-time case are given by the $D$-dimensional Poincar\'e algebra $\mf{iso}(D-1,1)$, with\footnote{Lorentz vector indices are $a=0,\ldots, D-1$ and we use the Minkowski metric $\eta_{ab} = \text{diag}(-++\ldots +)$ to raise and lower indices.}
\begin{align}
\label{eq:Poin}
\lb J_{ab} , J_{cd} \rb = 2 \eta_{c[b} J_{a]d} - 2\eta_{d[b} J_{a]c}\,,\quad\quad
\lb J_{ab}, P_c \rb = 2\eta_{c[b} P_{a]}\,,\quad\quad
\lb P_a, P_b \rb = 0\,,
\end{align}
We will take this as our starting point, and in order to allow for non-commuting translations we will introduce the most general extension of the Poincar\'e algebra.

Such extensions are classified by Lie algebra cohomology, and the most general possibility for the \Poincare~case is \cite{Bonanos:2008ez}
\begin{align}
\label{eq:Max2}
\lb P_a, P_b \rb = Z_{ab}\,,
\end{align}
where the Lorentz tensor $Z_{ab}$ is only constrained to be antisymmetric but is otherwise completely arbitrary. Note that this is a non-central extension due to the non-trivial Lorentz character of $Z_{ab}$. At this level we have also $[Z_{ab}, P_c]=0$.
This extension of the Poincar\'e algebra is known as the Maxwell algebra, and arises as the symmetry of a charged particle moving in a constant electro-magnetic background $F_{ab}$~\cite{Schrader:1972zd}, see also~\cite{Bacry:1968zf,Bacry:1970ye}.  

The (A)dS algebra~\eqref{eq:AdS} also has non-commuting translations. Due to the similarity with~\eqref{eq:Max2} one might be tempted to consider the identification of $Z_{ab} = J_{ab}$  as a quotient of Maxwell leading to the AdS isometry ($\sigma=+1$).\footnote{\label{fn:cx}As we will see in more detail below, the dS case is related to taking a different real slice of the complexified Maxwell algebra. This slice corresponds to taking $P_a\to i P_a$ which changes the sign in~\eqref{eq:Max2}.}
However, this kind of identification is inconsistent at this point: the relation $Z_{ab} -  J_{ab}$ that we wish to set to zero does not form an ideal of the algebra (as it commutes with $P$ into $P$) and therefore forces additional relations and the quotient by these relations is not the AdS algebra. 

There is, however, an approximate sense in which this is true. When including the relevant length scales, $P$ has the dimension of inverse length, and $Z$ therefore scales as $1/\LL^2$. 
The aformentioned commutator between $Z_{ab} -  J_{ab}$ and $P_c$ therefore only generates AdS generators up to order $1/\LL^3$ generators (as we will show in the next paragraph), and is therefore an approximate quotient. This suggests that one can continue the extension process to asymptotically attain (A)dS as a quotient.

Following this route and systematically including higher-order corrections to the Poincar\'e algebra, one can relax the condition that $[Z_{ab}, P_c]=0$ and include two further extensions here:
 \begin{align}
 \lb Z_{ab} , P_c \rb &=   {Y}_{ab,c} + 2 \eta_{c[b} Y_{a]} \,.
 \end{align}
We will use irreducible Lorentz representations throughout; hence $Y_{ab,c}=Y_{[ab],c}$ is traceless ($\eta^{bc} Y_{ab,c}=0$) and has mixed symmetry ($Y_{[ab,c]}=0$).

A putative AdS quotient at this stage would be to identify $Z_{ab} =  J_{ab}$ and $Y_a =  P_a$, and to set $Y_{ab,c}$ equal to zero. The commutator of translations with  $Z_{ab} - J_{ab}$ now generates  $Y_a - P_a$, which is also quotiented out. However, commuting translations with  $Y_a -P_a$ itself gives rise to $Z_{ab}$, and thus the three truncated generators do not form an ideal. In line with the previous discussion, however, the problem now only arises at order $1/\LL^4$, while the previous $1/\LL^3$ problem was solved by the introduction of the $Y$s. This suggests that continuation ad infinitum does allow for an AdS quotient.

Indeed such a structure is algebraically possible at higher order, with more and more irreducible representations of the Lorentz algebra added, triggering an infinite sequence of extensions~\cite{Bonanos:2008ez}. It was shown in~\cite{Gomis:2017cmt} that repeating this process ad infinitum produces the semi-direct product of the Lorentz algebra with the free Lie algebra generated by the translations $P_a$. We shall refer to the resulting algebra as $\Minfty$ as in~\cite{Gomis:2017cmt}. Moreover, the extension process naturally suggests the notion of `level' $\ell$ that counts how many $P_a$ one has to commute in order to produce the generator on the right-hand side. In this sense the generator $Z_{ab}$ of~\eqref{eq:Max2} is $\ell=2$. Higher levels can be generated using the techniques reviewed in~\cite{Gomis:2017cmt}. We shall also use the convention that $\ell=0$ contains the Lorentz generators $J_{ab}$. The first few levels are summarised in table~\ref{tab:Minf}. We shall denote the algebra obtained by retaining all generators on levels up to and including $\ell$ by $\text{Maxwell}_\ell$.

The notion of level agrees with the scaling behaviour of generators that we have discussed above. A generator at level $\ell$ scales as $1/\LL^\ell$ and therefore it is reasonable that, in order to have an exact expression such as the full (A)dS algebra, one has to go to infinite level.

\begin{table}[t!]
\centering
\begin{tabular}{c|cccccc}
level & $\ell=0$& $\ell=1$ & $\ell=2$ & $\ell=3$ & $\ell=4$ & $\cdots$\\\hline\\[-2mm]
$\mf{so}(D-1,1)$ irreps  & \raisebox{0.2\height}{\scalebox{0.6}{$\yng(1,1)$}}  & \raisebox{0.2\height}{\scalebox{0.6}{$\yng(1)$}}  & \raisebox{0.2\height}{\scalebox{0.6}{$\yng(1,1)$}}  &   \raisebox{0.2\height}{\scalebox{0.6}{$\yng(1)$}}   & \raisebox{0.2\height}{\scalebox{0.6}{$\yng(1,1)$}} ~ \raisebox{0.2\height}{\scalebox{0.6}{${\yng(2,1,1)}$}} & \\[2mm]
&&&&\raisebox{0.2\height}{\scalebox{0.6}{${\yng(2,1)}$}} &  \raisebox{0.2\height}{\scalebox{0.6}{$\yng(1,1)$}} ~ \raisebox{0.2\height}{\scalebox{0.6}{${\yng(2)}$}} \\
&&&&& $\bullet$   ~ \raisebox{0.2\height}{\scalebox{0.6}{${\yng(3,1)}$}} & $\cdots$
\end{tabular}
\caption{\label{tab:Minf}\it 
The Young tableaux of irreducible Lorentz representations that appear at the lowest levels $0\leq \ell \leq 4$ of $\Minfty$. The $\bullet$ represents the trivial representation of the Lorentz algebra, and the tableaux represent fully traceless representations.
}
\end{table}

It was shown in~\cite{Gomis:2017cmt}, extending results of~\cite{Bonanos:2008ez},  that the resulting infinite-dimensional Lie algebra $\Minfty$ admits a quotient to another infinite-dimensional Lie algebra that captures the Taylor expansion of the Lorentz equation of a charged particle in an 
arbitrary, $x$-dependent electro-magnetic background field that is not necessarily constant. This extends earlier work on the further truncation to the algebra only involving $\ell\leq 2$, i.e. stopping with the commutator~\eqref{eq:Max2}, that corresponds to restricting to constant electro-magnetic fields~\cite{Schrader:1972zd,Bacry:1970ye}.

\subsection{Quotient to $\Pinfty$}

Besides the electro-magnetic quotient, the algebra $\Minfty$ admits another quotient \cite{Gomis:2019fdh} that amounts to keeping only a subset of the Lorentz irreducibles shown in table~\ref{tab:Minf}. These Lorentz irreducibles consist of an infinite alternating sequence of vector and antisymmetric generators that we shall denote by
\begin{align}\label{quotientbn}
J_{ab}^{(m)} \quad (m\geq 0) \quad \text{and} \quad P_a^{(m)} \quad (m\geq 0)
\end{align}
where the $J_{ab}^{(m)}$ are the generators arising for even $\ell$ in table~\ref{tab:Minf} while the $P_a^{(m)}$ are the ones with odd $\ell$. Moreover, we have length dimensions 
\begin{align}
\label{eq:dimGen}
\left[ J_{ab}^{(m)} \right] = L^{-2m}\,,\quad\quad
\left[ P_a^{(m)} \right] = L^{-2m-1}\,,
\end{align}
for the different generators.

In order to cover both the dS and the AdS case, we complexify the Lie algebra spanned by these generators and consider two different real forms of the complex algebra. Our convention for AdS is such that $J_{ab}^{(0)}=J_{ab}$ and $P_a^{(0)} = P_a$. For the next levels this means $J_{ab}^{(1)} = Z_{ab}$ and $P_a^{(1)} = Y_a$ etc. For the dS case we rotate all the odd levels by the imaginary unit $i$ relative to the AdS case, i.e., we take $P_a^{(m)} \to i P_a^{(m)}$, see also footnote~\ref{fn:cx}. This turns out to yield a different real form of the same complexified algebra based on the generators~\eqref{quotientbn}.

The commutation relations in both cases can be determined because this quotient is also related to an affine Kac--Moody algebra~\cite{Gomis:2019fdh} and they read\footnote{We emphasise that the `mode numbers' in~\eqref{eq:AdSinf} are only non-negative and so this represents more precisely a parabolic subalgebra of a loop algebra. The central extension and the derivation operator (measuring the mode number) of affine Kac--Moody algebras can be consistently quotiented out in this parabolic subalgebra and are therefore not considered here.}
\begin{align}
\label{eq:AdSinf}
\lb J_{ab}^{(m)} , J_{cd}^{(n)}  \rb &= 2\eta_{c[b}^{\ } J_{a]d}^{(m+n)}- 2\eta_{d[b}^{\ } J_{a]c}^{(m+n)} \,,\nn\\
\lb J_{ab}^{(m)} , P_c^{(n)} \rb &= 2 \eta_{c[b}^{\ } P_{a]}^{(m+n)} \,,\\
\lb P_a^{(m)} , P_b^{(n)} \rb &= \sigma J_{ab}^{(m+n+1)}\,.\nn
\end{align}
This algebra has appeared in~\cite{Gomis:2019fdh} and in a non-relativistic split in~\cite{Khasanov:2011jr,Hansen:2019vqf}. We shall call it the $\AdSinfty$ algebra as it represents an infinite extension of the Poincar\'e algebra~\eqref{eq:Poin}.

We shall group together all even levels $\mf{g}_{2m}$, consisting of the generators $J_{ab}^{(m)}$ and call them the {\em generalised Lorentz algebra}\footnote{Using the same notion as here, the corresponding algebra in~\cite{Gomis:2019sqv} would be called the generalised homogeneous Galilei algebra.}
\begin{align}\label{generalizedLorentz}
\mf{L}_\infty = \bigoplus_{n=0}^\infty \mf{g}_{2n}\,,
\end{align}
while the odd levels will be referred to as {\em generalised translations}
\begin{align}
\mf{T}_\infty = \bigoplus_{n=0}^\infty \mf{g}_{2n+1}\,.
\end{align}
As a vector space decomposition we therefore have 
\begin{align}
\label{eq:symspace}
\AdSinfty= \mf{L}_\infty\oplus \mf{T}_\infty\,.
\end{align}
However, this is not a direct sum of Lie algebras as is clear from~\eqref{eq:AdSinf}: The generalised Lorentz algebra $\mf{L}_\infty$ is a closed Lie algebra that acts in a linear representation on the generalised translations $\mf{T}_\infty$. The generalised translations $\mf{T}_\infty$ do not commute and give back generalised Lorentz transformations. The algebraic structure here is that of a symmetric space decomposition.

One can truncate this infinite-dimensional algebra at a finite level by considering suitable quotients.  The simplest quotient of $\AdSinfty$ is obtained by considering only the generators $J_{ab}^{(0)}$ and $P_a^{(0)}$ while setting all generators with superscripts $>0$ equal to zero. This then produces an algebra isomorphic to the standard Poincar\'e algebra~\eqref{eq:Poin}. The next level would be Maxwell$_2$ with an additional $J_{ab}^{(1)}$ generator, while at $\ell=3$ there is another translation etc.

Another quotient, that will be more relevant in this paper, was alluded to before and requires the infinite-dimensional extension~\eqref{eq:AdSinf} of \Poincare; it fails to be a quotient when performed at finite order. The algebra $\Pinfty$ has an ideal that is  generated by 
\begin{align}
  P_a^{(0)} -P_a^{(m)} \,, \qquad J_{ab}^{(0)} - J_{ab}^{(m)} \,, \qquad \forall m>0 \,. \label{quotient}
 \end{align} 
The resulting quotient is isomorphic to the (A)dS given in~\eqref{eq:AdS}, with $\sigma=+1$ is related to AdS and $\sigma=-1$ to dS.\footnote{\label{footnote}Alternatively, one can obtain AdS by quotienting out by the ideal $P_a^{(0)} + (-)^{m+1} P_a^{(m)}$ and  $J_{ab}^{(0)} + (-)^{m+1} J_{ab}^{(m)}$ of the $\sigma = -1$ case of $\Pinfty$ (and similarly, a dS quotient from the $\sigma = +1$ case). However, these are related to the identifications in the main text by redefinitions, and hence will not be separately considered.}

The commutation relations (\ref{eq:AdSinf}) can be written in dual form in terms of the 
Maurer--Cartan one form $\Omega$
\begin{equation}
\label{MCforms}
\Omega=g^{-1} dg = \sum_{m=0}^\infty \left(e^a _{(m)}P_a ^{(m)}+\frac{1}{2}\omega^{ab} _{(m)}J_{ab} ^{(m)}\right)\,.
\end{equation}
Here, $g$ is a formal element of the the group obtained by exponentiating $\Pinfty$.
The components of $\Omega$ satisfy the Maurer--Cartan equations 
\begin{align}
de_{(m)}^a+\sum_{\substack{n,k\geq0\\n+k=m}}\omega^a_{(n)b}\wedge e_{(k)}^b&=0\,,\label{MCeqem}
\\
d\omega_{(m)}^{ab}+
\sum_{\substack{n,k\geq0\\n+k=m}}
\omega^{a}_{(n)c}\wedge \omega_{(k)}^{cb}
+\sigma \hskip-.4truecm\sum_{\substack{n,k\geq0\\n+k+1=m}}  e_{(n)}^a \wedge e_{(k)}^b&=0 \,,
\end{align}
which are a dual version of (\ref{eq:AdSinf}).

\subsection{Expansion of (A)dS}

As we will now demonstrate, the $\Pinfty$ algebra~\eqref{eq:AdSinf} is isomorphic to the algebra obtained by applying the Lie algebra expansion 
method~\cite{Hatsuda:2001pp,
deAzcarraga:2002xi,Izaurieta:2006zz}
with an infinite semigroup \cite{Penafiel:2016ufo} to the (A)dS algebra. 
We shall   consider the expansion method in the case when starting from a Lie algebra $\mf{g}$ that possesses a $\ints_2$-grading. By this we mean a decomposition $\mf{g}= \mf{g}_0 \oplus \mf{g}_1$ such that $[\mf{g}_i,\mf{g}_j]\subset \mf{g}_{i+j}$, 
 where the subscripts are taken modulo two.\footnote{A given algebra $\mf{g}$ can admit several such gradings and more general situations with larger gradings can also be considered~\cite{deAzcarraga:2002xi,Izaurieta:2006zz}.}
This is the case for the (A)dS algebra~\eqref{eq:AdS} with $\mf{g}_0=\langle J_{ab} \rangle$ and $\mf{g}_1= \langle P_a \rangle$. An infinite expansion can then be obtained by considering a graded Lie algebra with\footnote{This is the infinite-dimensional generalisation of the $\mathfrak{B}_{n}$ algebras introduced in~\cite{Izaurieta:2009hz} (see also \cite{Salgado:2014qqa,Concha:2014zsa}).}
\begin{align}
\label{eq:LAE}
\mf{g}_{2m} = \mf{g}_0 \otimes \LL^{-2m}\,,\qquad
\mf{g}_{2m+1} &= \mf{g}_1 \otimes \LL^{-2m-1}\,,
\end{align}
where we have taken the tensor product with the ring of power series $\cx[[\LL^{-1}]]$. The commutators of the expanded algebra then act by the usual commutator on the first factor and by the abelian product in the power series ring. The commutators in this graded algebra obey
\begin{align}
\lb \mf{g}_i , \mf{g}_j \rb \subset \mf{g}_{i+j} \quad\quad (i,j\geq 0).
\end{align}
We shall denote the generators of the even and odd level spaces as follows
\begin{align}
\mf{g}_{2m} : J_{ab}^{(m)}\equiv J_{ab} \otimes \LL^{-2m}  \quad\quad\text{and}\quad\quad \mf{g}_{2m+1} : P_a^{(m)}\equiv P_a\otimes \LL^{-2m-1}\,.
\end{align}
The algebra that they satisfy agrees with the $\AdSinfty$ algebra in~\eqref{eq:AdSinf} that arises as a quotient of $\Minfty$, which is why we have used the same notation. We thus find an equivalence between the bottom-up extension of \Poincare~and the top-down expansion of (A)dS\footnote{It follows from the quotient \eqref{quotient} that the expansion of AdS has an AdS quotient; amusingly, footnote \ref{footnote} implies that the expansion of AdS also has a dS quotient.}, and this is one of the key points of this paper. 

The above equivalence allows for a simple derivation of many properties of the $\Pinfty$ algebra. An example is the classification of invariant tensors. Given the (A)dS invariant bilinear form \eqref{AdS-bilinear-form},
the expansion method \cite{Izaurieta:2006zz} can be used to the define an invariant tensor on $\AdSinfty$ as
\begin{equation}\label{expinvt}
\begin{aligned}
&\langle J^{(m)}_{ab} , J^{(n)}_{cd}\rangle=\langle R^{-2m}\otimes J_{ab} ,R^{-2n}\otimes J_{cd}\rangle = \mu_{n+m}R^{-2(m+n)}\left(\eta_{ad}\eta_{bc}-\eta_{ac}\eta_{bd}\right) \,, \\
&\langle P^{(m)}_{a}, P^{(n)}_{b}\rangle=\langle R^{-2m-1}\otimes  P_{a}, R^{-2n-1}\otimes  P_{b}\rangle=\sigma\mu_{n+m+1}R^{-2(m+n+1)} \eta_{ab} \,,
\end{aligned}
\end{equation}
where $\mu_m$ stands for an infinite set of real dimensionless constants. The free coefficients $\mu_m$ correspond to the freedom of rescaling the $P_a^{(m)}$ and $J_{ab}^{(m)}$ for each $m$ separately. 
That this exhausts all choices of invariant bilinear forms on $\Pinfty$ can be seen by making a general ansatz  $\langle J_{ab}^{(m)}, J_{cd}^{(n)} \rangle = \left(\eta_{ad}\eta_{bc}-\eta_{ac}\eta_{bd}\right)  \alpha_{m,n}$ and $\langle P_a^{(m)} , P_b^{(n)} \rangle = \eta_{ab} \beta_{m,n}$ with symmetric $\alpha_{m,n}$ and $\beta_{m,n}$. The Lorentz structure is determined by invariance under $J_{ab}^{(0)}$ and evaluating invariance under the remaining generators leads quickly to the form~\eqref{expinvt}. We also note that the most general tensor on the space of translations $P_a^{(m)}$  that is invariant only under the generalised Lorentz algebra agrees with the second line of~\eqref{expinvt} by a similar argument.

\section{The $\Pinfty$ coset} 
\label{sec:NRPinfty}

In this section we consider a certain coset of $\Pinfty$ and its geometry in relation to Minkowski and (A)dS space.

\subsection{Non-linear realisation}

The $\Pinfty$ algebra represents an infinite generalisation of the Lorentz and translation generators. It is therefore natural to consider a non-linear realisation~\cite{Coleman:1969sm,Callan:1969sn,Salam:1969rq,Salam:1970qk,Isham:1971dv,Volkov:1973vd,Ogievetsky:1973ik,Ogievetsky:1974,Borisov:1974bn,Ivanov:1975zq}  of $\AdSinfty$ where the local subgroup is generated by all $J_{ab}^{(m)}$, which corresponds to the maximal subalgebra in the symmetric space decomposition~\eqref{eq:symspace}. The corresponding coset space will be denoted by 
$\STinfty$ and is associated with the remaining generalised translation generators $P_a^{(m)}$ for all $m\geq 0$ and a representative  can be written locally as
\begin{align}
\label{eq:Crep}
g = \exp \left( \sum_{m=0}^\infty x^a_{(m)} P_a^{(m)} \right) \,,
\end{align}
by choosing a particular gauge for the local generalised Lorentz generators $J_{ab}^{(m)}$. In view of  the dimensions listed in~\eqref{eq:dimGen}, we see that the coordinates $x^a_{(m)}$ on $\STinfty$ have dimensions $[x^a_{(m)}] = L^{2m+1}$ in order to make the argument of the exponential map dimensionless.

The non-linear transformations of the individual coordinates under the generalised translations follow in the standard way from the coset construction. We do this by considering left multiplication by $g_0$ and computing the effect infinitesimal transformation induced on the coordinates via $g_0 g h^{-1} = \exp \big( \sum_{m=0}^\infty (x^a_{(m)}+\delta x^a_{(m)}) P_a^{(m)} \big)$, where $h^{-1}$ is a generalised Lorentz transformation necessary to restore the gauge~\eqref{eq:Crep}. For generalised translations $g_0= \exp \big( \sum_{m=0}^\infty \epsilon^a_{(m)} P_a^{(m)} \big)$ evaluating this leads to the following relation between $\epsilon^a_{(m)}$ and $\delta x^a_{(m)}$ for any $m$:
\begin{align}
\label{eq:infcond}
& \delta x_{(m)}^a + \sum_{k\geq 1} \frac{\sigma^k}{(2k+1)!} \sum_{\substack{m_0,\ldots, m_{2k}\geq 0\\ m_0+\ldots m_{2k} +k=m} }\Big[ (x_{(m_{2k})}\cdot x_{(m_{2k-1})}) \cdots (x_{(m_{2})} \cdot x_{(m_{1})}) \delta x^a_{(m_0)} \nn\\
&\hspace{3.8cm} - (\delta x_{(m_{2k})}\cdot x_{(m_{2k-1})})( x_{(m_{2k-2})}\cdot x_{(m_{2k-3})}) \cdots (x_{(m_{2})} \cdot  x_{(m_{1})}) x^{a}_{(m_{0})} \Big]
\nn\\[8pt]
&= \epsilon^a_{(m)}+ \sum_{k\geq 1} \frac{\sigma^k}{(2k)!} \sum_{\substack{m_0,\ldots, m_{2k}\geq 0\\ m_0+\ldots m_{2k} +k=m} }\Big[ (x_{(m_{2k})}\cdot x_{(m_{2k-1})}) \cdots (x_{(m_{2})} \cdot x_{(m_{1})}) \epsilon^a_{(m_0)} \nn\\
&\hspace{4cm} - (\epsilon_{(m_{2k})}\cdot x_{(m_{2k-1})})( x_{(m_{2k-2})}\cdot x_{(m_{2k-3})}) \cdots (x_{(m_{2})} \cdot  x_{(m_{1})}) x^{a}_{(m_{0})} \Big]\,.
\end{align}
where $x\cdot y= x^a\eta_{ab} y^b$. Introducing the  non-normalised quadratic projection operators
\begin{equation}\label{Pop}
\overset{(mn)}{\mathcal Q^a{}_b}=x_{(m)}\cdot x_{(n)}\delta^a_b-
x^a_{(m)}x_{b(n)} \,,
\end{equation}
the solution to the~\eqref{eq:infcond} can be written as
\begin{align}
\label{eq:deltaxm}
&\delta x^a_{(m)} = \epsilon^a_{(m)} +\sum_{k\geq1} \frac{(4\sigma)^k B_{2k}}{(2k)!}
\hskip-.3truecm\sum_{\substack{m_0,\ldots m_{2k}\geq0\\m_0+\ldots+m_{2k}+k=m}}
\hskip-.2truecm
\overset{(m_{2k} m_{2k-1})}{\mathcal  Q^{a}_{\;\;b_{k-1}}}{\mathcal Q^{b_{k-1}}_{\;\;b_{k-2}}} 
\cdots
\overset{(m_{2} m_{1})}{\mathcal  Q^{b_{1}}_{\;\;b_{0}}}
\epsilon^{b_0}_{(m_{0})}\,.
\end{align}
in terms of the even Bernoulli numbers.\footnote{The Bernoulli numbers are given by the generating series $\frac{t}{e^t-1} = \sum_{n\geq 0} B_n \frac{t^n}{n!}$. From this one can deduce the formula $B_n=\sum\limits_{k=0}^n\sum\limits_{\ell=0}^k (-1)^\ell\binom{k}{\ell} \frac{\ell^n}{k+1}$. } This solution can be checked order-by-order. We shall present an all-order argument in section~\ref{sec:rel} when we consider the connection to (A)dS space. 

The calculation of the effect of an infinitesimal generalised Lorentz transformations is simpler since they act linearly on the generalised coordinates with the explicit form
 \begin{align}
 &\delta x^a_{(m)} = \sum_{\substack{m_1,m_2\geq 0\\m_1+m_2=m}} \epsilon^{ab}_{(m_1)} \eta_{bc} x_{(m_2)}^c\,,
\end{align}
with parameter $\epsilon^{ab}_{(m)}$. We shall see again that this transformation can be related to (A)dS space in section~\ref{sec:rel}.

\subsection{Invariant metrics}

In order to construct invariant metrics for the coset $\STinfty$, we turn to the Maurer--Cartan forms~\eqref{MCforms} and evaluate them in the parametrisation~\eqref{eq:Crep}. 
Evaluating the vielbein for $\STinfty$ leads to
\begin{equation}
\label{eq:CMe}
e^a_{(m)} = dx^a_{(m)}+\sum_{k\geq 1} \frac{\sigma^k}{(2k+1)!} \hskip-.2truecm
\sum_{\substack{m_0,\ldots, m_{2k}\geq 0\\ m_0+\ldots +m_{2k} +k=m} }
\hskip-.2truecm
\overset{(m_{2k} m_{2k-1})}{\mathcal  Q^{a}_{\;\;b_{k-1}}}{\mathcal Q^{b_{k-1}}_{\;\;b_{k-2}}} 
\cdots
\overset{(m_{2} m_{1})}{\mathcal  Q^{b_{1}}_{\;\;b_{0}}}
dx^{b_0}_{(m_{0})} 
\end{equation}
using the non-normalised projectors \eqref{Pop}.
The second terms in~\eqref{MCforms} are the composite spin connections arising from the generalised Lorentz symmetry; in complete analogy to the \Poincare~case these are not independent but can be expressed in terms of the coordinates of translations. Explicitly, one finds
 \begin{align}\label{omegaMCform}
\omega^{ab}_{(m)} &= - 2\sum_{k\geq 0} \frac{\sigma^{k+1}}{(2k+2)!} \hskip-.2truecm\sum_{\substack{m_0,\ldots, m_{2k+1} \geq 0\\ m_0+\ldots+m_{2k+1} + k +1= m}} \hskip-.5truecm \left(x_{(m_{2k+1})}\cdot x_{(m_{2k})}\right) \cdots \left(x_{(m_{3})} \cdot x_{(m_{2})}\right) x_{(m_{1})}^{[a} dx_{(m_0)}^{b]} \,.
\end{align}

The Maurer--Cartan forms transform linearly under generalised Lorentz transformations according to 
 \begin{align}
\label{transformaitonstheta}
 \delta e^a _{(m)}= \sum_{\substack{n,k\geq0\\n+k=m}}\theta^{ab}_{(n)}\eta_{bc} e^c _{(k)}\,.
 \end{align}
 This formula is true in general and $\theta^{ab}_{(n)}$ is associated with the compensating generalised Lorentz transformation of the non-linear realisation. For generalised translations this means that it depends on the coordinates $x^a_{(m)}$.

The Maurer--Cartan forms $e^a_{(m)}$ can be used to construct an invariant metric for the coset space $\STinfty= \exp(\AdSinfty)/\exp(\mf{L}_\infty )$ as\footnote{We have included the factor $R^2$ in front of the invariant line element to obtain the Minkowski metric as the leading term both in the dS and the AdS case.} 
\begin{equation}\label{lineelementPinfty}
ds^2=\sigma R^2 \sum_{m,n=0}^{\infty}  e_{(m)}^a e_{(n)}^b\langle P_a^{(m)}, P_b^{(n)} \rangle \,,
\end{equation}
where $\langle\,,\rangle$ us an invariant metric whose most general form was given in~\eqref{expinvt}. 
It can be expanded as
\begin{equation}
\label{lineelementPinfty2}
ds^2= \sum_{m=0}^\infty \mu_m R^{-2m} ds^2_{(m)}
\,,
\hskip1truecm ds^2_{(m)}= \sum_{\substack{n,k\geq0\\n+k=m}} e_{(n)} \cdot e_{(k)}\,.
\end{equation}
where the line elements $ds_{(m)}^2$ can be worked out as
\begin{equation}\label{subL}
\begin{aligned}
ds^2_{(0)}&= e_{(0)}^2=\eta_{ab}dx^a_{(0)} dx^b_{(0)}\,,\\[6pt]
ds^2_{(1)}&= 2e_{(0)}\cdot e_{(1)}= 2\eta_{ab}dx^a_{(0)} dx^b_{(1)}
+\frac{\sigma}{3}\left(x_{(0)}^{2}\eta_{ab}-x_{a(0)}x_{b(0)} \right) dx^a_{(0)} dx^b_{(0)}\,, \\[6pt]
ds^2_{(2)}&= e_{(1)}^2+2e_{(0)}\cdot e_{(2)}\\
&= \eta_{ab}\left(2dx^a_{(0)} dx^b_{(2)} +dx^a_{(1)} dx^b_{(1)}\right)
+\frac{2\sigma}{3}\left(x_{(0)}^{2}\eta_{ab}
-x_{a(0)}x_{b(0)} \right) dx^a_{(0)} dx^b_{(1)}\\
&\quad +\frac{2\sigma}{3}\left( x_{(0)} \cdot x_{(1)} \eta_{ab} -x_{a(0)}x_{b(1)}\right) dx^a_{(0)} dx^b_{(0)}
+
\frac{2x_{(0)}^2}{45}\left(x_{(0)}^{2}\eta_{ab}-x_{a(0)}x_{b(0)} \right) dx^a_{(0)} dx^b_{(0)}\,.
 \\
\vdots
\end{aligned}
\end{equation}
It is useful to note that every line element can be written as
\begin{equation}
ds^2_{(m)}=g^{(m)}_{AB}dX_{(m)}^A dX_{(m)}^B\,,
\end{equation}
where $g^{(m)}_{AB}$ is a $\big((m+1)D\big)$-dimensional metric tensor with coordinates $X^A_{(m)}=\{x^a_{(0)}\,\dots,x^a_{(m)}\}$ The metrics $g^{(m)}_{AB}$ are non-degenerate and define geometries with vanishing Ricci scalar,  with non-vanshing Riemann and Ricci tensors. 
Given a finite truncation of $\AdSinfty$ spanned by generators $P_a^{(0)},\dots,P_a^{(N-1)}$, and $J_{ab}^{(0)},\dots,J_{ab}^{(N)}$, an invariant metric $\langle P_a^{(m)}, P_b^{(n)} \rangle$ for the translation generators can be obtained from \eqref{expinvt} by setting $\mu_{m> N}=0$, and it is non-degenerate for $\mu_N\neq0$.

\subsection{Relation to Minkowski and (A)dS space}
\label{sec:rel}

We now discuss how the infinite-dimensional generalised space $\STinfty$ with coordinates $x^a_{(m)}$ given in~\eqref{eq:Crep} relates to known $D$-dimensional space-times, in a spirit similar to~\cite{Gomis:2019sqv}. The basic idea is to define a family of hypersurfaces of co-dimension $D$ by imposing $D$ linear relations on the coordinates $x^a_{(m)}$. A point of the $D$-dimensional space-time is then identified with a full hypersurface and the symmetries of the $D$-dimensional space-time relate different hypersurfaces to each other. In particular, this means that motion within the hypersurface is invisible to an observer who has only access to the $D$-dimensional space-time which is a quotient space of the infinite-dimensional space $\STinfty$ by the hypersurface.

The first, somewhat trivial, example of such a family of hypersurfaces  we consider is
\begin{align}
\label{eq:Mink1}
z^a = x^a_{(0)}\,,
\end{align}
so that the hypersurfaces are labelled only by the coordinate $x^a_{(0)}$ and the values of the $x^a_{(m)}$ with $m>0$ can take any value and are unconstrained. Note that $z^a$ has dimension of length according to dimensions of the $x^a_{(m)}$ discussed below~\eqref{eq:Crep}.

Let us consider how the symmetries~\eqref{eq:deltaxm} act on the $z^a$. Due to~\eqref{eq:Mink1} we only need to consider what happens to $x^a_{(0)}$, the motion within the hypersurface corresponding to the other coordinates $x^a_{(m)}$ for $m>0$ is irrelevant. Clearly,~\eqref{eq:deltaxm} reduces to
\begin{align}
\delta z^a = \epsilon^a_{(0)} + \epsilon^{ab}_{(0)} z_b\,,
\end{align}
exactly the usual Poincar\'e transformations, telling us that~\eqref{eq:Mink1} corresponds to identifying Minkowski space as a quotient of the infinite-dimensional $\STinfty$. Since none of the higher generators of $\AdSinfty$ have an effect on $z^a$, we can also consider the quotient algebra of $\AdSinfty$ by all generators $P^{(m)}_a$ and $J^{(m)}_{ab}$ for $m>0$. This quotient is isomorphic to the usual Poincar\'e algebra.

\medskip

The second example of a family of hypersurfaces that we consider is defined by the relation\footnote{This expansion is similar to what was done in~\cite{VandenBleeken:2017rij,Hansen:2019vqf,Gomis:2019sqv,Ergen:2020yop} in a $1/c$ expansion for non-relativistic systems. }
\begin{align}
\label{eq:coordexp1}
x^a = \sum_{m=0}^\infty \LL^{-2m-1} x^a_{(m)}\,.
\end{align}
Note that this $x^a$ is dimensionless. 
We can again consider the effect of a transformation~\eqref{eq:deltaxm} on $x^a$, leading to 
\begin{equation}
\label{eq:deltax2}
\delta x^a =\big[\delta^a_b+\left( r  \coth r-1\right) \mathcal P^a{}_{b}\big] \epsilon^b   + \epsilon^{ab} x_b\,,
\end{equation}
where $r^2 = \sigma x^a \eta_{ab} x^b$ and 
\begin{align}
\label{eq:epscoll}
\epsilon^a = \sum_{m\geq 0} \LL^{-2m-1} \epsilon^a_{(m)}\,,\qquad 
\epsilon^{ab} = \sum_{m\geq 0} \LL^{-2m} \epsilon^{ab}_{(m)}
\end{align}
are collective parameters formed out of the individual parameters appearing in~\eqref{eq:deltaxm}. We emphasise that the transformation~\eqref{eq:deltax2} is valid for any choice of individual parameters $\epsilon^a_{(m)}$ when they are combined into the collective expressions~\eqref{eq:epscoll}. Here, we have also introduced the projector
\begin{equation}
\mathcal P^a{}_b=\delta^a_b -\frac{x^a x_b}{x^2}\,,
\hskip1.5truecm  
\mathcal P^a_{\;\;c} \mathcal P^c_{\;\;b}=\mathcal P^a_{\;\;b} \,.
\end{equation}
The transformation agrees precisely with the transformation~\eqref{eq:deltaxAdS} of the coordinates of (A)dS space in the coordinates introduced in appendix~\ref{app:AdS}. As shown there, the translation part of the transformation~\eqref{eq:deltax2} is determined by the condition \eqref{eq:AdScond1} 
\begin{align}
\label{eq:AdScondd}
\big[\delta^a_b+\left(\cosh r-1\right) \mathcal P^a_{\;\;b} \big] \epsilon^b  =\left[ \delta^a_b + \left(\frac{\sinh r -r}{r}\right) \mathcal P^a_{\;\;b}   \right] \delta x^b\,.
\end{align}
Expanding this condition using~\eqref{eq:coordexp1} and~\eqref{eq:epscoll} as well as
\begin{align}
\label{eq:r2exp}
r^{2k} &= \sigma^k \left[\sum_{m,n=0}^\infty \LL^{-2(m+n)} x_{(m)}{\cdot}x_{(n)}\right]^k \!\!
= \sigma^k \!\!\!\!\!\!  \sum_{m_1,\ldots,m_{2k}=0}^\infty \!\!\!\! \LL^{-2\sum_{i=1}^{2k} m_i} (x_{(m_1)}{\cdot} x_{(m_2)}) \cdots (x_{(m_{2k-1})}{\cdot}x_{(m_{2k})}) \,,
\end{align}
one recovers from this condition the equation~\eqref{eq:infcond}. As~\eqref{eq:AdScondd} determines~\eqref{eq:deltax2}, the solution to the expanded version~\eqref{eq:infcond} of~\eqref{eq:AdScondd} can be determined by expanding~\eqref{eq:deltax2}. Doing this and using
\begin{align}
\label{eq:expHyp}
r \coth r = \sum_{k=0}^\infty  \frac{4^k B_{2k}}{(2k)!}  r^{2k} \,,
\end{align}
we deduce~\eqref{eq:deltaxm}.  This shows that the non-linear realisation of $\AdSinfty$ is linked to the symmetries of (A)dS space if one makes the formal expansion~\eqref{eq:coordexp1} and~\eqref{eq:epscoll}. This situation is similar to the one in~\cite{Gomis:2019sqv} where the $1/c$-expansion was considered except for the fact that now the translations are non-commuting and the transformation laws are thus non-linear.

As the collective parameter~\eqref{eq:epscoll} still has many individual parameters $\epsilon^a_{(m)}$ it appears as if the symmetry of (A)dS space is enlarged from the usual (A)dS algebra to an infinite-dimensional algebra. However, the effective symmetry group acting on the quotient space is not bigger than the (A)dS algebra since the effective transformation~\eqref{eq:deltax2} on the collective coordinate is exactly that of (A)dS space.

Similarly, by applying \eqref{eq:coordexp1} in the (A)dS metric \eqref{eq:AdSmetric}
\begin{equation}
ds^2_{\rm (A)dS}=g_{ab}dx^a dx^b
\,,\hskip1truecm
g_{ab}=\eta_{ab}+\left(\frac{\sinh^2 r}{r^2}-1\right) \mathcal P_{ab} \,,
\end{equation}
leads to the $\AdSinfty$ invariant metric \eqref{lineelementPinfty2}.

{Equations~\eqref{eq:Mink1} and~\eqref{eq:coordexp1} represent only two possible choices of hypersurfaces for which we have an immediate physical interpretation. Clearly, the space $\STinfty$ admits many other choices in the same way that the algebra $\Pinfty$ also has many other possible quotients. It would be interesting to find other meaningful examples.}

\section{The $\Pinfty$ particle}
\label{sec:part}

In order to construct a particle action that is invariant under the $\AdSinfty$ algebra using the method of non-linear realisation~\cite{Coleman:1969sm,Callan:1969sn,Salam:1969rq,Salam:1970qk,Isham:1971dv,Volkov:1973vd,Ogievetsky:1973ik,Ogievetsky:1974,Borisov:1974bn,Ivanov:1975zq}, we will consider the most general quadratic expression involving bilinear invariant tensors\footnote{One could construct more general actions using higher-order invariant tensors, but we expect these to be subleading in an effective field theory expansion.}; in other words, we base our particle action on the line element \eqref{lineelementPinfty}:
\begin{equation}\label{actionPinfty}
S
= \frac{ \sigma R^2 m}{2}\int d\tau\, \sum_{m,n=0}^{\infty} e_{(m)}^a e_{(n)}^b\langle P_a^{(m)}, P_b^{(n)} \rangle \,,
\end{equation}
where $\langle \;,\; \rangle$ stands for an invariant bilinear form~\eqref{expinvt} on the $\AdSinfty$ algebra. Here, the pullback of the Maurer--Cartan forms for the generalised translations to the particle worldline is understood and therefore we implement $dx^\mu _{(m)}\rightarrow \dot x^\mu _{(m)}$ in the Maurer--Cartan forms $e^a_{(m)}$. If we restrict our analysis to Lagrangians that are quadratic in the Maurer--Cartan forms, the result \eqref{actionPinfty} is the most general action that is invariant under the generalised Lorentz algebra \eqref{generalizedLorentz}.

An expansion of this action leads to the following expressions at lowest order:
\begin{equation}\label{actionsum}
S=\sum_{m=0}^\infty  \mu_m R^{-2m} S_m\,,
\hskip1truecm S_{m}=\frac{m}{2}\int d\tau \sum_{\substack{n,k\geq0\\n+k=m}} e_{(n)} \cdot e_{(k)}\,,
\end{equation}
where
\begin{equation}\label{subS}
\begin{aligned}
S_0&
= \frac{ m}{2}\int d\tau\, \dot x_{(0)}^2\,,\\[6pt]
S_1&
= m
\int d\tau\left[\dot{x}_{(0)}\cdot\dot{x}_{(1)}+\frac{\sigma}{3!}\left(x_{(0)}^{2}\dot{x}_{(0)}^{2}-\left(\dot{x}_{(0)}\cdot x_{(0)}\right)^{2}\right)\right]   \,,\\[6pt]
S_2
&= m\int d\tau\Bigg[
\frac{1}{2}\dot{x}_{(1)}^{2}+\dot{x}_{(0)}\cdot\dot{x}_{(2)}-\frac{\sigma}{3}\dot{x}_{(0)}\cdot x_{(0)}\left(\dot{x}_{(1)}\cdot x_{(0)}+\dot{x}_{(0)}\cdot x_{(1)}\right)\\
&\hskip1.5truecm+\frac{\sigma}{3}\left(x_{(0)}^{2}\dot{x}_{(1)}\cdot\dot{x}_{(0)}+x_{(0)}\cdot x_{(1)}\dot{x}_{(0)}^{2}\right)+
\frac{x_{(0)}^2}{45}\left(x_{(0)}^{2}\dot{x}_{(0)}^{2}-\left(x_{(0)}\cdot \dot x_{(0)}\right)^{2}\right)
\Bigg]
 \\
\vdots
\end{aligned}
\end{equation}
Note that the expression at each order is fully fixed by generalised Lorentz invariance. There are however free overall coefficients $\mu_m$ at every order. 

When turning to the equations of motion following from the full action, it turns out that the values of these coefficients are completely irrelevant when we consider the summed action up to some finite order $N$. The reason is that there is a telescopic (or Matryoshka) structure to the equations of motion following from~\eqref{actionsum}. Referring back to~\eqref{eq:CMe} we see that $x_{(N)}^a$ only occurs in $e_{(m)}^a$ for $m\geq N$ and therefore only via $e_{(N)}^a$ when the sum in~\eqref{actionsum} is truncated at $N$. In $S_N$ it occurs only in the form $S_N \propto\int \dot{x}_{(N)} \cdot \dot{x}_{(0)}+\ldots$ and therefore 
\begin{align}
\frac{\delta S_N}{\delta x_{(N)}^a} \propto \ddot{x}_{(0)}^a
\end{align}
enforces the equation of motion $\ddot{x}^a_{(0)}=0$ irrespective of the value of $\mu_N$ (as long as it is non-zero which we assume without loss of generality). When next computing the variation with respect to $x_{(N-1)}^a$ there are contributions from both $S_{N-1}$ and $S_N$:
\begin{align}
\frac{\delta S}{\delta x_{(N-1)}^a} = \mu_N R^{-2N} \frac{\delta S_N}{\delta x_{(N-1)}^a} + \mu_{N-1} R^{-2(N-1)}\underbrace{ \frac{\delta S_{N-1}}{\delta x_{(N-1)}^a} }_{\propto \,\ddot{x}^a_{(0)}=0}  \approx \mu_N R^{-2N} \frac{\delta S_N}{\delta x_{(N-1)}^a}
\end{align}
and therefore the contribution from the next lower equation of motion vanishes on-shell, such that the last equality only holds on-shell. Therefore the value $\mu_{N-1}$ is irrelevant on-shell and this implies that one can compute the full equations of motion simply from the highest action $S_N$ occurring in a given truncation. 

Due to the above structure, the field equations take a universal form (independent of the coefficients $\mu_m$): 
\begin{align}
\label{eqsS2}
&\delta x_{(N)}^{a}:\hskip.4truecm&
\ddot{x}_{(0)}^a&= 0\,,\nn\\[6pt]
&\delta x_{(N-1)}^{a}:\hskip.4truecm&
\ddot{x}_{(1)}^{a}&= \frac{2\sigma}{3}\left(\dot{x}_{(0)}^{2}x_{(0)}^{a}-\dot{x}_{(0)}\cdot x_{(0)}\dot{x}_{(0)}^{a}\right)
\,\nn\\[6pt]
&\delta x_{(N-2)}^{a}:\hskip.4truecm&
\ddot{x}_{(2)}^{a}&=\frac{2\sigma}{3}\left(\dot{x}_{(0)}^{2}x_{(1)}^{a}+2\dot{x}_{(1)}\cdot\dot{x}_{(0)}x_{(0)}^{a}-x_{(0)}\cdot\dot{x}_{(0)}\dot{x}_{(1)}^{a} -\frac{d}{d\tau}\left(x_{(0)}\cdot x_{(1)}\right)\dot{x}_{(0)}^{a}\right)\nn\\
&&&\quad +\frac{2}{45}\left(x_{(0)}^{2}\dot{x}_{(0)}\cdot x_{(0)}\dot{x}_{(0)}^{a}+3x_{(0)}^{2}\dot{x}_{(0)}^{2}x_{(0)}^{a}-4\left(\dot{x}_{(0)}\cdot x_{(0)}\right)^{2}x_{(0)}^{a}\right)\,.\\
&\vdots&&\nn
\end{align}

Since we are considering one-dimensional world-lines to explore the geometry of $\STinfty$ and the same geometry is encoded in the Maurer--Cartan equations~\eqref{MCeqem}, we can also summarise the equations of motion in these terms.
Pulling back the Maurer--Cartan forms for $e_{(m)}^a$ and $\omega^{ab}_{(m)}$ to the world-line by replacing $dx^\mu _{(m)}\rightarrow \dot x^\mu _{(m)}$ everywhere, the Maurer--Cartan equation implies
\begin{equation}
\dot e_{(m)}^a+\sum_{\substack{n,k\geq0\\n+k=m}}\omega^a_{(n)b}e_{(k)}^b=0\,.
\end{equation}
This equation is completely equivalent to~\eqref{eqsS2} when truncated at level $N$.

The above equation should be seen as the geodesic motion through the higher-dimensional space-time spanned by all generalised coordinates. As we pointed out in section 3.3, there is a linear combination of these (consisting of their weighted sum  \eqref{eq:coordexp1}) that transforms as AdS coordinates under $\Pinfty$ transformations. What dynamics is implied by the $\Pinfty$ particle for this particular set of coordinates? The equations \eqref{eqsS2} can be put together into one single equation for $x^a$ that reads
\begin{equation}
\label{geodesic0}
\ddot{x}^{a}=\frac{2\sigma}{3 R ^{2}}\left(\dot{x}^{2}x^{a}-x\cdot\dot{x}\dot{x}^{a}\right)+\frac{2}{45 R ^{4}}\left(x^{2}x\cdot\dot{x}\dot{x}^{a}+3x^{2}\dot{x}^{2}x^{a}-4\left(x\cdot\dot{x}\right)^{2}x^{a}\right)+\dots
\end{equation}
As shown in appendix~\ref{app:part} this is precisely the expansion of a particle moving in AdS space. This underlines our perspective on the role of the higher-order generators: including these (and their associated coordinates) allows one to iteratively build up the AdS geodesic equation, including more and more curvature corrections as one goes further up in level. 

At the first non-trivial level, one can see the known effect that AdS space acts like a confining box by studying~\eqref{geodesic0} up to order $\LL^{-2}$.  The first term on the right-hand side is transverse to $\dot{x}^a$ as can be checked easily. The leading order of the resulting  motion can be written as
\begin{align}\label{tmotion}
\mathcal{T}^a{}_b \ddot{x}^b = \frac{2\sigma \dot{x}^2}{3\LL^2} \mathcal{T}^a{}_b x^b+ \mathcal{O}(\LL^{-4})\,, \quad \big(\delta^a_b- \mathcal{T}^a{}_b\big) \ddot{x}^b = 0 + \mathcal{O}(\LL^{-4})\,,
\end{align}
where $\mathcal{T}^a{}_b=\delta^a{}_b - \dot{x}^{-2} \dot{x}^a\dot{x}_b$ is the projector transverse to the velocity.
 For affine parametrisation $\dot{x}^2$ is a negative constant and in AdS space ($\sigma=+1$) this means that the transverse motion is that of a harmonic oscillator.  
The longitudinal motion is free for both AdS and dS to this order since 
$ \ddot{x}\cdot \dot{x}=0$. 
 For the case of dS 
 we have an inverted harmonic oscillator and
the symmetries of the transverse motion
are the Newton--Hooke symmetries \cite{Bacry:1968zf}.
 One can continue and include the $R^{-4}$ order, which can be put in the form
\begin{align}
 \ddot{x}^a = \frac{2\sigma \dot{x}^2}{3\LL^2} \mathcal{T}^a{}_b x^b
+\frac{2}{45\LL^4} \dot{x}^2 x^2 \left(3\mathcal T^a{}_b -4 \mathcal P^a{}_c \mathcal T^c{}_b \right)x^b +\mathcal{O}(R^{-6})\,,
\end{align}
and also includes non-trivial longitudinal motion.

Note that one can also consider alternatives to the quadratic action introduced above; using linearly the pullback of the MC form as in e.g.~\cite{Gomis:2017cmt}, one could also consider a reparametrisation invariant particle action of the form
\begin{equation}\label{actionPinftyRI}
\tilde S =- m R^2 \int d\tau\, \Bigg[ -\sigma\sum_{m,n=0}^{\infty} e_{(m)}^a e_{(n)}^b\langle P_a^{(m)}, P_b^{(n)} \rangle \Bigg]^{1/2}\,,
\end{equation}
which corresponds to the usual action of degree one in the derivatives for a relativistic particle $\int ds$. In this case, using the invariant bilinear form \eqref{lineelementPinfty} tensor allows one to Taylor expand the square root in powers of $R^{-2}$ for $R>1$. With a suitably defined notion of proper time, we expect that this action in proper time gauge leads to equivalent dynamics as the quadratic one that we have studied above.

\section{Conclusions and Outlook}

In this paper we have shown how
the (A)dS corrections to the Poincar\'e algebra can be naturally and perturbatively
described within the free Maxwell algebra ($\Minfty$) approach. 
This algebra was introduced in~\cite{Gomis:2017cmt} in order to describe
the most general coupling of particles to an electromagnetic
general background that in general could depend on multipoles.
A quotient of this algebra 
 gives the unfolded description of the Maxwell field
\cite{Vasiliev:2005zu,Boulanger:2015mka} and
captures the Taylor expansion of the Lorentz equation of a charged particle in an $x$-dependent
arbitrary electro-magnetic background field that is not necessarily constant.  

Instead, in this paper we have discussed a different quotient of $\Minfty$ that contains only
vectors and two-forms. 
It represents an infinite extension of the Poincar\'e algebra, termed the $\AdSinfty$ algebra~\eqref{eq:AdSinf}, with the vector space decomposition  
\begin{align}
\AdSinfty= \mf{L}_\infty\oplus \mf{T}_\infty \,,
\end{align}
that is an indecomposable representation of the generalised Lorentz algebra $\mf{L}_\infty$. By considering a non-linear realisation of $\AdSinfty$ with respect to $\mf{L}_\infty$ we end up with an infinite-dimensional space $\STinfty$ with generalised coordinates $x^a_{(m)}$ ($m\geq 0$). 
The transformations of $\AdSinfty$ on these generalised coordinates can be computed from the non-linear realisation. Defining a set of hypersurfaces of $\STinfty$  by 
\begin{align}
x^a = \sum_{m=0}^\infty \LL^{-2m-1} x^a_{(m)}\,, \label{hyper}
\end{align}
we showed that $\AdSinfty$ transformations on $x^a$ are the (A)dS transformations. We have also constructed the most general particle action that is quadratic in derivatives~\eqref{actionsum}, which was shown to generate the expansion in $1/R$ of the geodesic equation in an (A)dS background~\eqref{actionpartAdS} (when projected onto the hypersurface \eqref{hyper}). It follows from our results that the same algebraic structure $\Minfty$ allows for quotients and non-linear realisations that play a role in different physical contexts, ranging from the Lorentz force in flat space-time to geodesic motion in curved space-times like (A)dS. 

One of the mathematical properties that allows one to identify $\Pinfty$ with (A)dS curvature corrections is the isomophism between the bottom-up free Lie algebra extension of \Poincare{} and the isometry algebra expansion of AdS. Starting from the \Poincare{} algebra in the Levi decomposition,  this implies that it is equivalent to extend the non-semi-simple algebra or to first deform the algebra into the simple AdS algebra and then to expand this algebra. 

It appears to us that the equivalence between these two operations transcends the particular case of this paper. For instance, the same perspective holds for the non-relativistic Galilei algebra~\cite{Gomis:2019fdh}. In this case, the free Lie algebra has a quotient that can be called Galilei$_{\infty}$. Similarly, one can deform the Galilei algebra into the \Poincare{} algebra and subsequently expand this algebra. This should result in the same Galilei$_{\infty}$ algebra and amounts to including relativistic $1/c$ corrections in two different but equivalent ways. It would be interesting to consider other contexts where these algebraic procedures can also shed light on physical theories and limits thereof. An example might be the Carrollian limit of relativistic theories, corresponding to $c \rightarrow 0$. 

A useful tool for writing the full algebra~\eqref{eq:AdSinf} was that it coincides with a parabolic subalgebra of an affine Kac--Moody algebra~\cite{Gomis:2019fdh} which is clear from the equivalence to the expansion method. At the level of the free Lie algebra $\Minfty$ this means that the ideal that is quotiented out corresponds to certain Serre relations. It would be interesting to consider other similar quotients, leading potentially to non-affine but more general Kac--Moody algebras and see what their corresponding physics is. Moreover, it is intriguing that only the parabolic half of a Kac--Moody algebra appears since all mode numbers are positive and may wonder whether the negative levels can be included in a meaningful way.

\subsection*{Acknowledgments}

We are grateful to Sander Andela, Andrea Barducci,
Roberto Casalbuoni, Dijs de Neeling, Tom\'{a}s Ort\'{\i}n, Patricio Salgado, Tonnis ter Veldhuis and Jelmar de Vries for stimulating discussions. JG   thanks the Van Swinderen Institute for Particle Physics and Gravity,
University of Groningen, for its hospitality and creative
atmosphere.
DR also thanks the Universitat de Barcelona for its hospitality and stimulating atmosphere.  

JG has been supported in part by MINECO FPA2016-76005-C2-1-P and Consolider CPAN, and by the Spanish government (MINECO/FEDER) under project MDM-2014-0369 of ICCUB (Unidad de Excelencia Mara de Maeztu).
PS-R acknowledges the School of Physics and Astronomy of the University of Leeds for hospitality and support as invited researcher.

\appendix

\section{The (Anti) de Sitter case}
\label{app:AdS}

In this appendix, we collect a number of relevant known aspects of the (Anti) de Sitter case, including its algebra, non-linear realisation and particle action.

\subsection{The   algebra}

The AdS algebra $\mf{so}(D-1,2)$ for $\sigma=+1$ and the dS algebra $\mf{so}(D,1)$ for $\sigma=-1$  in $D$ space-time dimensions are given by
\begin{align}
\label{eq:AdS}
\lb J_{ab} , J_{cd} \rb = 2 \eta_{c[b} J_{a]d} - 2\eta_{d[b} J_{a]c}\,,\quad\quad
\lb J_{ab}, P_c \rb = 2\eta_{c[b} P_{a]}\,,\quad\quad
\lb P_a, P_b \rb = \sigma 
J_{ab}\,.
\end{align}
Here, we have absorbed the length scale $\LL$ of (A)dS space into the translation generators which are therefore dimensionless. Here, $J_{ab}$ are the $\mf{so}(D-1,1)$ Lorentz generators and $P_a$ are the dimensionless (A)dS translations that do not commute.

The (A)dS algebra can also be written in dual form using the Maurer--Cartan form 
\begin{align}
\Omega =  g^{-1} dg= E^a P_a + \frac12 \Omega^{ab} J_{ab}\,,
\end{align}
where $g$ is an element of the (A)dS group, via the Maurer--Cartan equations
\begin{align}
\label{eq:AdSMCE}
dE^a +\Omega^a{}_b E^b=0 \,, \qquad
d\Omega^{ab} +\Omega^{a}{}_c \Omega^{cb}+\sigma E^{a}{E}^b&=0\,.
\end{align}

The invariant bilinear form for this algebra is given by
\begin{equation}
\langle J_{ab}, J_{cd}\rangle= \eta_{ad}\eta_{bc}-\eta_{ac}\eta_{bd} \,, \hskip.7truecm \langle P_{a}, P_{b}\rangle= \sigma\eta_{ab}\,.
\label{AdS-bilinear-form}
\end{equation}

\subsection{The  (A)dS coset}
\label{relationAdS}

We consider the non-linear realisation of the (A)dS algebra~\eqref{eq:AdS} with respect to the Lorentz algebra, leading to (A)dS space-time.
The coset representative is given by  
\begin{align}\label{parametrizationgAdS}
g = \exp\big( x^a P_a \big) \,,
\end{align}
and the coordinates are also dimensionless. The (A)dS scale $\LL$ can be introduced later by rescaling both the generators and the coordinates. 

The left-invariant Maurer--Cartan form 
in this parametrisation of the coset
can be calculated as 
\begin{align}
E^a &=dx^a
+\left(\frac{\sinh r}{r}-1\right)\mathcal P^{a}_{\;\;b}dx^b\,, \qquad 
\Omega^{ab} = -2 (\cosh r-1) \frac{x^{[a} dx^{b]}}{x^2}\,,
\end{align}
where $r^2=\sigma x^a\eta_{ab} x^b=\sigma x^2$.
From~\eqref{eq:AdSMCE} evaluated for the coset~\eqref{parametrizationgAdS} we see that (A)dS  space-time has a constant curvature.
 
The (A)dS metric in these coordinates takes the form 
\begin{align}
\label{eq:AdSmetric}
ds^2 = E^a \eta_{ab} E^b =dx^a \eta_{ab} dx^b+ \left(\frac{\sinh^2 r}{r^2}-1\right) dx^a  \mathcal P_{ab} dx^b  \,.
\end{align}
Depending on the sign of $\sigma$ the hyperbolic trigonometric functions can become ordinary trigonometric functions for the dS case.\footnote{ 
This coordinate system is not global as can be seen by embedding these coordinates into an ambient space according to 
\begin{align}
u^a=\frac{x^a}{r} \sinh r\,,  \qquad
u^\sharp = \cosh r \,, \nn
\end{align}
where the (A)dS space is defined as the following hypersurface in the ambient space
\begin{align}
\eta_{ab}u^a u^b-\sigma (u^\sharp)^2&= - \sigma \,,\nn
\end{align}
This coordinate system is not global in AdS since $|u^\sharp|\geq 1$ for $\sigma=+1$. }

In this coordinate system, the algebra~\eqref{eq:AdS} is realised by the dimensionless vector fields 
\begin{align}
\label{eq:AdSvecs}
J_{ab} &= x_a \partial_b - x_b \partial_a \,, \qquad 
P_a = \left[ \delta^b_a + \left(r  \coth r-1\right) \mathcal P_{a}^{\;\;b}\right] \partial_b \,,
\end{align}
that generate the infinitesimal transformations 
\begin{align}
\label{eq:deltaxAdS}
\delta x^a =\left[\delta^a_b+\left( r  \coth r-1\right)\mathcal P^a_{\;\;b}\right] \epsilon^b   + \epsilon^{ab} x_b
\end{align}
on the coordinates $x^a$. Here, we have given $P_a$ the parameter $\epsilon^a$ and $J_{ab}$ the parameter $\frac12 \epsilon^{ab}$. This transformation is the same we have deduced from $\STinfty$ in~\eqref{eq:deltax2} above.

We note that the length scale $\LL$ of (A)dS space can be reintroduced by letting the coordinates $x^a \to  \LL^{-1} x^a$ and parameter $\epsilon^a \to  \LL^{-1} \epsilon^a$ so that the new coordinates and parameters become dimensionful with $[x^a]=L^1$ and $[\epsilon^a]=L^1$. The transformation~\eqref{eq:deltaxAdS} then reads 
\begin{align}
\label{eq:deltayAdS}
\delta x^a =\left[\delta^a_b+\left( \frac{r}{R}  \coth (r/R)-1\right) \mathcal P^a_{\;\;b}\right] \epsilon^b   + \epsilon^{ab} x_b\,,
\end{align}
where  $r^2 =\sigma x^2$ is now also dimensionful, so that the combination $r/\LL$ is dimensionless. Taking the limit $r/\LL\to 0$ reduces the symmetries to the Poincar\'e symmetries of flat space.

The condition that determines the translation part of the transformation~\eqref{eq:deltaxAdS} in the non-linear realisation is 
\begin{align}
\label{eq:AdScond1}
\left[\delta^a_b+\left(\cosh r-1\right) \mathcal P^a_{\;\;b} \right] \epsilon^b  =\left[ \delta^a_b + \left(\frac{\sinh r}{r}-1\right) \mathcal P^a_{\;\;b}    \right] \delta x^b
\end{align}
and this condition is used in section~\ref{sec:rel} to connect to the non-linear realisation of $\AdSinfty$. 

\subsection{The   particle}
\label{app:part}

The action of a massive particle moving on (A)dS space reads
\begin{equation}\label{actionpartAdS}
\begin{aligned}
S&=\frac{m}{2}\int d\tau \,g_{ab}\dot x^a \dot x^b \,,
\end{aligned}
\end{equation}
where the metric $g_{ab}$ is given in \eqref{eq:AdSmetric}. The corresponding geodesic equation is
\begin{equation}\label{geqAdS}
\ddot{x}^a+\Gamma^{a}_{bc}\dot{x}^{b}\dot{x}^{c}=0 \,,
\end{equation}
where $\Gamma^{a}_{bc}$ are the torsion free
Christoffel symbols 
$\Gamma^{a}_{bc}=\frac{1}{2}g^{ad}\left(\partial_{b}g_{dc}+\partial_{c}g_{db}-\partial_{d}g_{bc}\right)$.
Using the parametrisation \eqref{parametrizationgAdS} and reinserting the (A)dS radius according to section~\ref{relationAdS}, the  (A)dS metric \eqref{eq:AdSmetric} and its inverse take the form  
\begin{align}
g_{ab}&= \eta_{ab}+\left(\frac{R^2}{r^2}\sinh^2\left(\frac{r}{R}\right)-1\right) \mathcal P_{ab} \,, \qquad
g^{ab}=\eta^{ab}+\left(\frac{r^2}{R^2}{\rm csch}^2\left(\frac{r}{R}\right) -1\right)\mathcal P^{ab} \,.
\end{align}
Replacing this back in \eqref{geqAdS} leads to 
\begin{align}\label{geqAdS2}
\ddot{x}^{a}&=-2\left(\frac{r}{R}\coth\frac{r}{R}-1\right)\frac{x\cdot\dot{x}}{r^2}\dot{x}^{a}-\left(1-\frac{R}{r}\sinh\frac{r}{R}\cosh\frac{r}{R}\right)\frac{\dot{x}^{2}}{r^2}x^{a}\\[6pt]
&\hskip.5truecm-\left(1+\frac{R}{r}\sinh\frac{r}{R}\cosh\frac{r}{R}-\frac{2r}{R}\coth\frac{r}{R}\right)\frac{\left(x\cdot\dot{x}\right)^{2}}{r^{4}}x^{a}\nn\\
&=\frac{2\sigma}{3 R ^{2}}\left(\dot{x}^{2}x^{a}-x\cdot\dot{x}\dot{x}^{a}\right)+\frac{2}{45 R ^{4}}\left(x^{2}x\cdot\dot{x}\dot{x}^{a}+3x^{2}\dot{x}^{2}x^{a}-4\left(x\cdot\dot{x}\right)^{2}x^{a}\right)+\dots\,.
\end{align}
The expansion in powers of $1/R$ agrees precisely with \eqref{geodesic0}.

\end{document}